\documentclass[aps,twocolumn,prl,showpacs]{revtex4}%
\usepackage{bm}
\usepackage{mathrsfs}
\usepackage{amsmath}
\usepackage{amssymb}
\usepackage{graphicx}
\usepackage{amsfonts}
\usepackage{amsthm}
\usepackage{color}
\usepackage{dcolumn}
\usepackage{txfonts}

\begin{document}
\title{Universality of Uhrig dynamical decoupling for suppressing qubit pure dephasing and relaxation}
\author{Wen Yang}
\author{Ren-Bao Liu}
\thanks{rbliu@phy.cuhk.edu.hk}
\affiliation{Department of Physics, The Chinese University of Hong Kong, Shatin, N. T., Hong Kong, China}


\begin{abstract}
The optimal $N$-pulse dynamical decoupling discovered by
Uhrig for a spin-boson model [Phys. Rev. Lett, {\bf 98}, 100504 (2007)] is proved to
be universal in suppressing  to $O\left(T^{N+1}\right)$
the pure dephasing or the longitudinal relaxation of a qubit (or spin-1/2)
coupled to a generic bath in a short-time evolution of duration $T$.
It is also found that for the purpose of suppressing the longitudinal relaxation,
an ideal Uhrig $\pi$-pulse sequence can be generalized to a sequence consisting
of the ideal one superimposed with finite-duration pulses
satisfying certain symmetry requirements.
\end{abstract}

\maketitle

{\it Introduction} -- A central topic in spin resonance
spectroscopy~\cite{Slichter1992} is the decoherence of spins due to
coupling to environments, including the longitudinal relaxation of the
population and the transverse relaxation of the phase correlation
(i.e., dephasing) in the basis quantized along an external magnetic
field~\cite{Pines1955,Joos2003_Book}. Also, the decoherence of a
qubit, which can be modeled by a spin-1/2, is the main obstacle in
implementing scalable quantum computing~\cite{Nielsen2000_Book}. To
deal with the spin or qubit decoherence, various strategies have
been developed, including quantum error
correction~\cite{Shor:1995PRA,Steane:1996PRL,Knill:1997PRA,Zanardi:1997PRL},
decoherence-free subspace~\cite{Duan:1997PRL,Lidar:1998PRL},
dynamical decoupling (DD) or bang-bang
control~\cite{Mehring_NMR,Rhim1970_PRL,haeberlen1976hrn,
Viola1998_PRA,Viola1999_PRL,Viola2005_PRL,Kern2005_PRL,Khodjasteh2005_PRL,Khodjasteh:2007PRA,Santos:2006PRL,
Yao2007_RestoreCoherence,Witzel2007_PRBCDD,Zhang:2007PRB,Uhrig2007UDD},
and dynamical control by  pulse spectrum or shape engineering~\cite{Agarwal:1999PRA,Kofman:2001PRL,Gordon:2008PRL}.
In particular, the DD suppresses the decoherence by
eliminating the qubit-bath coupling through stroboscopic rotation of the qubit.
An especially interesting DD scheme is the concatenated
DD~\cite{Khodjasteh2005_PRL,Khodjasteh:2007PRA,Santos:2006PRL,
Yao2007_RestoreCoherence,Zhang:2007PRB,Witzel2007_PRBCDD} which
applies recursively a lower-order periodic pulse sequence as the
building block of the next higher order sequence.
For an evolution of a short duration $T$, an $N$th order
concatenated DD eliminates the qubit-bath coupling up to
$O\left(T^{N+1}\right)$. The number of pulses in concatenated DD,
however, increases exponentially with increasing order $N$. Since
errors are inherently introduced by the controlling pulses, it is
desirable to have DD sequences with the minimum number of
controlling pulses.

An optimal DD scheme was first discovered
by Uhrig for a pure dephasing spin-boson
model~\cite{Uhrig2007UDD}, which uses $N$ $\pi$-pulses applied at
\begin{equation}
T_j=T\sin^{2}\frac{j\pi}{2(N+1)},\ \ \ {\rm for}\ j=1,2,\ldots, N,
\label{UhrigTj}
\end{equation}
to eliminate the dephasing up to $O\left(T^{N+1}\right)$. Optimal
pulse sequences for $N\le 5$ have also been noticed by Dhar
\textit{et al.} earlier for controlling the Zeno
effect~\cite{Dhar2006_PRL}. Lee, Witzel and Das Sarma conjectured
that the Uhrig dynamical decoupling (UDD) may work for a generic
pure-dephasing model with an analytical verification up to
$N=9$~\cite{Lee2008_PRL}. Later computer-assisted algebra was
used to verify the conjecture up to $N=14$~\cite{UhrigArxiv2008}.
Aiming at a general proof of the conjecture, Cardy and
Dhar~\cite{CardyArxiv2008} have given a very inspiring though
unsuccessful attempt by formulating the problem in a time-dependent
perturbation theory.

In this Letter, we shall complete the proof of the universality of
the UDD in suppressing the pure dephasing or the longitudinal
relaxation of a qubit (or spin-1/2) coupled to a generic bath. The
proof is based on the observation that to preserve the spin
coherence up to a given order, one does not have to eliminate all
terms of the effective qubit-bath coupling to the given order as in
a generic concatenated DD but just needs to eliminate the terms
relevant to the decoherence. An extension of the
proof is that an ideal UDD sequence, for countering the
longitudinal spin relaxation, can be replaced with
a more general UDD sequence consisting of the same ideal
$\pi$-pulses and some extra finite-duration pulses as long as
certain symmetry requirements are fulfilled.

{\it Ideal UDD for a generic pure dephasing Hamiltonian} --
Let us first consider the ideal UDD pulse sequences for a Hamiltonian
of the form
\begin{align}
\hat{H}=\hat{C}+\hat{\sigma}_z\otimes\hat{Z},
\end{align}
where $\hat{\sigma_z}$ is the qubit Pauli matrix along the
$z$-direction, and $\hat{C}$ and $\hat{Z}$ are bath operators. This
Hamiltonian describes a pure dephasing model for it contains no
qubit flip processes and therefore leads to no longitudinal
relaxation but only transverse dephasing. Pure dephasing models are
of special interest in quantum computing since very often the qubit
flip terms can be essentially eliminated by applying a strong static
magnetic field along the $z$-direction. A specific example is the
spin-boson model in which
$\hat{C}=\sum_{i}\omega_{i}\hat{b}_{i}^{\dagger} \hat {b}_{i}$ and
$\hat{Z}=\sum_{i}\left( \lambda_{i}/2\right)
\left(\hat{b}_{i}^{\dagger}+\hat{b}_{i}\right)$ with $\hat{b}_{i}$
being a boson annihilation operator. It is for this spin-boson model
that Uhrig~\cite{Uhrig2007UDD} has discovered the optimal pulse sequences with
timing given in Eq.~(\ref{UhrigTj}).

Now we shall prove that the UDD applies for arbitrary $\hat{C}$ and $\hat{Z}$.
To overcome the pure dephasing, the ideal UDD sequences consist of
$\delta$-pulse $\pi$-rotations about a transverse axis (say, the $x$-axis)~\cite{Uhrig2007UDD}.
The qubit dephasing is characterized by the decay of the
expectation value of the raising or lowering operator $\hat{\sigma}_{\pm}\equiv
\hat{\sigma}_x\pm i\hat{\sigma}_{y}$,
\begin{align}
L_{+,-}(T)\equiv
\left|\left\langle\hat{\sigma}_{-}(T)\right\rangle\right|
=\left|\left\langle
\hat{\sigma}_{-}\hat{U}^{(N)\dag}_-\hat{U}^{(N)}_+
\right\rangle\right|,
\end{align}
where the qubit state-dependent bath propagators under the $N$th order UDD control are
\begin{align}
\hat{U}_{\pm}^{(N)}
=e^{-i\left[\hat{C}\pm (-1)^{N} \hat{Z}\right]\left(T-T_{N}\right)}
\cdots
e^{-i\left(\hat{C}\mp\hat{Z}\right)\left(T_{2}-T_{1}\right)}
e^{-i\left(\hat{C}\pm\hat{Z}\right)T_{1}}.
\label{Upm}
\end{align}
To show that the $N$th order UDD suppresses the pure dephasing up to $O\left(T^{N+1}\right)$ for
a small $T$, we just need to prove
\begin{align}
\hat{U}_{-}^{(N)\dagger}\hat{U}^{(N)}_{+}=1+O\left(  T^{N+1}\right).
\label{propagator}
\end{align}
By expanding the difference $\delta\hat{U} \equiv \hat{U}_{+}-\hat{U}_{-}$
into Taylor series and collecting the coefficients term by term,
Uhrig has verified Eq.~(\ref{propagator}) for $N\le 14$ with computer-assisted
algebra~\cite{UhrigArxiv2008}.

We shall proceed with the formalism of the time-dependent perturbation theory
due to Cardy and Dhar~\cite{CardyArxiv2008}. With the standard
formulation in the interaction picture, Eq. (\ref{Upm}) can be put
in the time-ordered formal expression
\begin{equation}
\hat{U}^{(N)}_{\pm}=e^{-i\hat{C}T}\hat{\mathscr T}e^{-i\int_{0}^{T}\pm
F_{N}\left(t\right) \hat{Z}_I\left( t\right)  dt},
\label{UpmTorder}%
\end{equation}
where $\hat{\mathscr T}$ is the time-ordering operator, the
modulation function $F_{N}\left(  t\right)\equiv
\left(-1\right)^{j}$ for $t\in \left[  T_{j},T_{j+1}\right]$ with
$T_0\equiv 0$ and $T_{N+1}\equiv T$, and
\begin{align}
\hat{Z}_{I}(t)  \equiv & e^{i\hat{C} t }\hat{Z}e^{-i\hat{C}t} =
\sum_{p=0}^{\infty}\frac{(it)^p}{p!}\underbrace{\left[\hat{C},
\left[\hat{C},\cdots\left[\hat{C}, \hat{Z} \right] \cdots\right]
\right]}_{p\ {\rm folds}} \equiv \sum_{p=0}^{\infty}\hat{Z}_pt^p.
\label{expansion}
\end{align}
The difference $\delta \hat{U}$ is given by the Taylor series%
\begin{subequations}
\begin{align}
\delta\hat{U}=2e^{-i\hat{C}T}
\sum_{k=0}^{\infty}\left(-i\right)^{2k+1} \hat{\Delta}_{2k+1},
\end{align}
with
\begin{align}
\hat{\Delta}_{n}\equiv &
\int_{0}^{T}F_{N}\left(t_{n}\right)
\int_{0}^{t_{n}}F_{N}\left( t_{n-1}\right)
\cdots
\int_{0}^{t_{2}}F_{N}\left(  t_{1}\right)
\nonumber \\
& \times\left[
 \hat{Z}_I\left(t_{n}\right) \hat{Z}_I\left(t_{n-1}\right) \cdots \hat
{Z}_I\left(t_{1}\right)\right]dt_{1}dt_{2}\cdots dt_{n}.
\end{align}
\end{subequations}
A feature of the expansion of the difference $\delta\hat{U}$ is that
it contains only odd-order terms $\hat{\Delta}_{2k+1}$ which are
relevant to the dephasing. We just need to show
$\hat{\Delta}_{2k+1}=O\left(T^{N+1}\right)$.

Using the expansion in Eq.~(\ref{expansion}), we have
\begin{align}
\hat{\Delta}_{n}    =  \sum_{\left \{  p_{j}\right \}  }
 \left[\hat{Z}_{p_n}\cdots \hat{Z}_{p_2}\hat{Z}_{p_1}
F_{p_1,p_2,\cdots,p_n} T^{n+p_{1}+p_2\cdots+p_{n}} \right ],
\label{TaylorSeries}
\end{align}
where
\begin{align}
 F_{p_1,\cdots,p_n}\equiv
\int_{0}^T \frac{dt_n}{T} \cdots \int_{0}^{t_3}
\frac{dt_2}{T}\int_{0}^{t_2} \frac{d t_1}{T} \prod_{j=1}^n
F_{N}\left(t_{j}\right) \left (\frac{t_j}{T}\right )^{p_j} \nonumber
\end{align}
is a dimensionless constant independent of $T$.

Now the problem is reduced to prove
\begin{equation}
F_{p_1,p_2,\cdots,p_n}=0
\label{PolyLemma}
\end{equation}
for $n$ being odd and $n+\sum_{j=1}^{n}p_{j}\le N$. For this
purpose, we make the variable substitution $t_{j}=T\sin^{2}
(\theta_{j}/2)$ and define the scaled modulation function
$f_{N}\left(\theta\right)\equiv
F_N\left(T\sin^{2}(\theta/2)\right)=\left(-1\right)^{j}$ for
$\theta\in\left[j\pi/(N+1),\left(j+1\right)\pi/(N+1)\right]$. With
$\sin^{2p}(\theta/2)\sin \theta=(2i)^{-2p}\sum_{r=0}^{2p} C_{2p}^{r}\sin
\left[\left(p-r+1\right)\theta\right]$, we can write
$F_{p_1,p_2,\cdots,p_n}$ as a linear combination of terms of the
form
\begin{align}
f_{q_1,\cdots,q_n} \equiv
\int_{0}^{\pi} d\theta_n\cdots \int_{0}^{\theta_{3}} d\theta_2  \int_{0}^{\theta_2} d\theta_1
\prod_{j=1}^{n}f_{N}\left(\theta_j\right)\sin\left(q_j\theta_j\right),
\nonumber
\end{align}
with $\left|q_{j}\right|\le p_j+1$. Suffices it to show
$f_{q_1,q_2,\cdots,q_n}=0$ for odd $n$ and $\sum_{j=1}^n\left|q_j\right|\le N$.
We notice that $f_{N}\left(\theta\right)$ has period of $2\pi/(N+1)$
and hence expand it into Fourier series
\begin{equation}
f_{N}\left(\theta\right)
=\sum_{k=1,3,5,\cdots}\frac{4}{k \pi } \sin \left[ k \left( N+1\right) \theta \right].
\label{fxFourier}
\end{equation}
The key feature of the Fourier expansion to be exploited is that it contains
only odd harmonics of $\sin[(N+1)\theta]$.
With the Fourier expansion, we just need to show that
\begin{equation}
\int_0^{\pi} d\theta_n \cdots  \int_{0}^{\theta_3} d\theta_2 \int_{0}^{\theta_2}d\theta_1
\prod_{j=1}^n\cos\left(r_j \theta_j+q_j\theta_j\right)=0,
\label{CosLemma0}
\end{equation}
for $n$ being odd, $r_j$ being an odd multiple of $(N+1)$,
and $\sum_{j=1}^n\left|q_j\right|\le N$.
With the product-to-sum trigonometric formula repeatedly used,
it can be shown by induction that after an even number
of variables $\theta_1, \theta_2, \dots, \theta_{2k}$
have been integrated over, the resultant integrand as a function of $\theta_{2k+1}$
can be written as the sum of cosine functions of the form
\begin{align}
\cos\left(R_{2k+1}\theta_{2k+1}+Q_{2k+1}\theta_{2k+1}\right),
\end{align}
with $R_{2k+1}$ being an odd multiple of $(N+1)$  and
$\left|Q_{2k+1}\right|\le \sum_{j=1}^{2k+1}\left|q_j\right|$.
In particular, the last step is
\begin{align}
\int_0^{\pi} \cos\left(R_n \theta_n +Q_n \theta_n \right)d\theta_n.
\label{CosLemma}
\end{align}
Since $R_n$ is an odd (non-zero, of course) multiple of $(N+1)$,
and $\left|Q\right|\le \sum_{j=1}^n \left| q_j\right|\le N$, we have
$R_n+Q_n \ne 0$ and the integral above must be zero.
Thus Eq.~(\ref{PolyLemma}) holds.
The proof is done.

{\it Ideal UDD for suppressing longitudinal spin relaxation} --
Now we consider the most generic qubit-bath Hamiltonian
\begin{equation}
\hat{H}=\hat{C}+\hat{\sigma}_{x}\otimes \hat{X}+\hat{\sigma}_{y}\otimes \hat{Y}+\hat{\sigma}_{z}\otimes \hat{Z},
\end{equation}
where $\hat{\sigma}_{i}$ are the Pauli matrices of the qubit and
$\hat{C}$, $\hat{X}$, $\hat{Y}$, and $\hat{Z}$ are bath operators.
Without loss of generality, we assume the $z$-axis as the rotation
axis for qubit control. We aim to show that the spin polarization
along the rotation axis $\left|\langle
\hat{\sigma_z}(T)\rangle\right|$ is preserved up to
$O\left(T^{N+1}\right)$ under the control of the $N$th order UDD.
The spin polarization is
\begin{align}
\left|\langle \hat{\sigma_z}(T)\rangle\right|=\left|\langle
\hat{U}^{(N)\dag}\hat{\sigma}_z \hat{U}^{(N)}\rangle\right|,
\end{align}
where the propagator $\hat{U}$ can be written as
\begin{align}
\hat{U}^{(N)}  =
e^{-i\left[\hat{C}'+(-1)^N\hat{D}\right](T-T_{N})}\cdots
e^{-i\left(\hat{C}'-\hat{D}\right)(T_2-T_{1})}
e^{-i\left(\hat{C}'+\hat{D}\right)T_1},
\label{UTorder}
\end{align}
in which the Hamiltonian has been separated into
$\hat{C}' \equiv \hat{C}+\hat{\sigma}_{z}\otimes \hat{Z}$ and
$\hat{D}\equiv \hat{\sigma}_{x}\otimes \hat{X}+\hat{\sigma}_{y}\otimes\hat{Y}$.
With the definition $\hat{D}_{I}\left( t\right)\equiv e^{i\hat{C}'t}\hat{D}e^{-i\hat{C}'t}$,
the propagator can be formally expressed as
\begin{align}
\hat{U}^{(N)}=e^{-i\hat{C}' T} \hat{\mathscr T} e^{-i\int_0^T F_N(t)\hat{D}_I(t) dt },
\end{align}
which has the same form as Eq. (\ref{UpmTorder}).
Following the same procedure as for proving Eq.~(\ref{Upm}), we find that
up to $O\left(T^{N+1}\right)$, the expansion of the propagator contains
only terms consisting of even power of $\hat{D}$.
Since $\hat{D}$ contains only the Pauli matrices $\hat{\sigma}_x$
and $\hat{\sigma}_y$ and an even power of the two Pauli matrices
$\hat{\sigma}_x^{n_x}\hat{\sigma}_y^{n_y}$ (with $n_x+n_y$ being even)
is either unity or $i \hat{\sigma}_z$, the propagator
\begin{align}
\hat{U}^{(N)}=e^{-i\hat{H}_{\rm eff}T+O\left(T^{N+1}\right)},
\end{align}
where the effective Hamiltonian $\hat{H}_{\rm eff}$ commutes with $\hat{\sigma}_z$.
Thus the $N$-pulse UDD eliminates the
longitudinal qubit relaxation up to $O\left(T^{N+1}\right)$.

\textit{UDD with non-ideal pulses: Longitudinal relaxation} -- With
the help of Eq.~(\ref{CosLemma0}), we realize that
Eq.~(\ref{PolyLemma}) holds for more general modulation functions
$F_N(t)$ as along as the scaled modulation function
$f_N(\theta)\equiv  F_{N}\left(T\sin^2(\theta/2)\right)$ contains
only odd harmonics of $\sin[(N+1)\theta]$ as in Eq.~(\ref{fxFourier}), i.e,
\begin{align}
f_N(\theta)=\sum_{k=0}^{\infty}A_k\sin\left [(2k+1)(N+1)\theta\right],
\label{odd_harmonics}
\end{align}
with arbitrary coefficients $A_k$. Motivated by this observation,
we try to generalize the UDD to the case of non-ideal pulses.

\begin{figure}[ptb]
\includegraphics[width=\columnwidth]{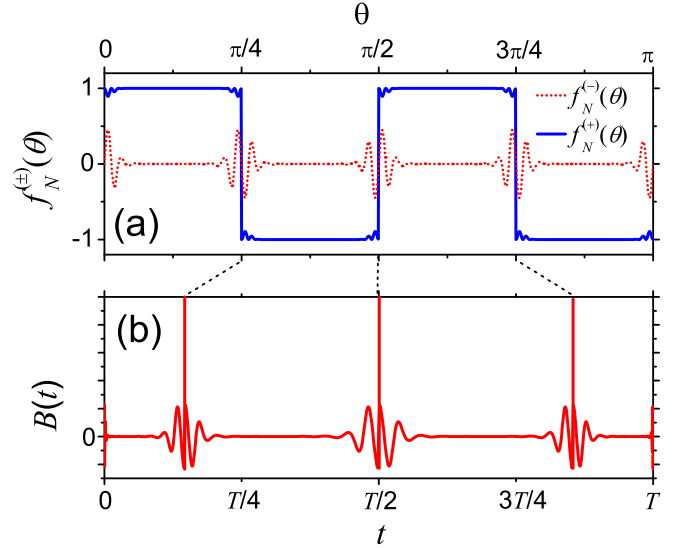}
\caption{(color online) An example of (a) the scaled modulation functions
$f_{N}^{\pm}(\theta)$ for the generalized 3rd order UDD control
and (b) the corresponding magnetic field $B(t)$. The dashed lines
indicate the correspondence between the sudden jumps of the
modulation function $f^{+}_N(\theta)$ in (a) and the sharp spikes as ideal
$\pi$-pulses in (b).}%
\label{Pulses}
\end{figure}

Consider the control of the qubit by an arbitrary time-dependent
magnetic field $B(t)$ applied along the $z$-direction, the general
qubit-bath Hamiltonian is
\begin{align}
\hat{H}(t)=\hat{C}+\hat{\sigma}_{x}\otimes\hat{X}+\hat{\sigma}_{y}\otimes \hat{Y}
 +\hat{\sigma}_{z}\otimes \hat{Z}+\frac{1}{2}\hat{\sigma}_{z}B(t).
\end{align}
In the rotating reference frame following the qubit precession
under the magnetic field, the Hamiltonian becomes
\begin{align}
\hat{H}_R(t)=\hat{C}'+\cos[\phi(t)]\hat{D}^{+}+\sin[\phi(t)]\hat{D}^{-},
\end{align}
where the precession angle $\phi(t)=\int_0^{t}B\left(t'\right)dt'$,
$\hat{C}'\equiv \hat{C}+\hat{\sigma}_{z}\otimes \hat{Z}$,
$\hat{D}^{+}\equiv \hat{\sigma}_{x}\otimes\hat{X}+
\hat{\sigma}_{y}\otimes \hat{Y}$, and $\hat{D}^{-}\equiv
\hat{\sigma}_{x}\otimes\hat{Y}- \hat{\sigma}_{y}\otimes \hat{X}$.
The propagator in the rotating reference frame is
\begin{align}
\hat{U}=e^{-i\hat{C}'T}\hat{\mathscr T}\exp \left(-i\int_0^{T}\sum_{\lambda=\pm}
F_{N}^{\lambda}(t) D_{I}^{\lambda}(t)dt\right),
\end{align}
with $F_{N}^{+}(t)=\cos [\phi(t)]$, $F_{N}^{-}(t)=\sin[\phi(t)]$,
and
$\hat{D}_{\mathrm{I}}^{\lambda}(t)=e^{i\hat{C}'t}\hat{D}^{\lambda}e^{-i\hat{C}'t}$.
To consider the qubit relaxation, we just need to examine the odd
power of $\hat{D}^{\pm}$ in the Taylor expansion of the propagator.
The same way as derive Eq.~(\ref{TaylorSeries}), we find that
for the $n$th power of $\hat{D}^{\pm}$, the expansion in $T$ has
coefficients as
\begin{align}
T^{n+p_1+p_1+\cdots+p_n} \int_{0}^T \frac{dt_n}{T} \cdots \int_{0}^{t_3}
\frac{dt_2}{T} \int_{0}^{t_2} \frac{dt_1}{T} \prod_{j=1}^n
F_{N}^{\lambda_j}\left(t_{j}\right)
\left(\frac{t_j}{T}\right)^{p_j}.
\nonumber
\end{align}
For $n$ being odd and $n+\sum_{j=1}^{n}p_{j}\le N$, the multiple
integral above vanishes [such that the qubit relaxation is
suppressed to $O\left(T^{N+1}\right)$] as long as the scaled
modulation function $f^{\pm}_N(\theta)\equiv
F^{\pm}_N\left(T\sin^{2}(\theta/2)\right)$ contains only odd harmonics of
$\sin[(N+1)\theta]$ as depicted in Eq.~(\ref{odd_harmonics}). This
condition is satisfied if and only if the scaled modulation functions $f_{N}^{\pm}(\theta)$
have the following symmetries:
\begin{enumerate}
\item periodic with period of $2\pi/(N+1)$;
\item anti-symmetric with respect to $\theta=j\pi/(N+1)$;
\item symmetric with respect to $\theta=(j+1/2)\pi/(N+1)$.
\end{enumerate}
The anti-symmetry condition requires $f_{N}^{\pm}(\theta)$
be either zero or discontinuous at $\theta=j\pi/(N+1)$. But $f_{N}^{+}(\theta)$
and  $f_{N}^{-}(\theta)$ cannot be simultaneously zero
since they have to satisfy the normalization condition
\begin{equation}
\left[ f_{N}^{+}(\theta)\right]^{2}+\left[f_{N}^{-}(\theta)\right]^{2}=1,
\label{fxNorm}
\end{equation}
according to the definition of $F^{\pm}_N(t)$.
So there must be sudden jumps at least in one of two modulation
functions at $\theta=j\pi/(N+1)$, which means the controlling
magnetic field $B(t)$ has to contain a $\delta$-pulse for
$\pi$-rotation at $t=T_j$. With the initial conditions
$f_{N}^{+}(0)=1$ and $f_{N}^{-}(0)=0$, one can choose the field
such that $f_{N}^{-}(\theta)$ is continuous while $f_{N}^{+}(0)$
has sudden jumps between $+1$ and $-1$ at $\theta=j\pi/(N+1)$.
Thus, a generalized UDD sequence can be chosen the following way:
For $\theta\in\left[0,\pi/(2N+2)\right]$, $f_{N}^{+}(\theta)$ can be
arbitrary but sudden jumps from $-1$ to $+1$ at $\theta=0$ and from
$+1$ to $-1$ at $\pi/(2N+2)$, and $f_{N}^{-}(\theta)$ is determined from
the normalization condition as
$f_{N}^{-}(\theta)=\pm \sqrt{1-\left[f^{+}_N(\theta)\right ]^{2}}$.
At other regions, $f_{N}^{\pm}(\theta)$ are determined by the symmetry requirements.
The pulse amplitude $B(t)$ for the generalized UDD is
\begin{align}
B(t)=\frac{1}{F^{+}_N(t)}\frac{d}{dt}{F}^{-}_N(t)=\sum_{j=1}^N\pi \delta\left(t-T_j\right)
+B_{\rm extra}(t),
\end{align}
which is a superposition of the ideal UDD pulses and an extra
component $B_{\rm extra}(t)$ being arbitrary but subject to
the symmetry requirements. The demand of $\delta$-pulses in the
generalized UDD is consistent with the previous finding in
Ref.~\cite{PasiniPRA2008} that the effect of an ideal $\pi$-pulse
on the evolution of a qubit coupled to a bath cannot be exactly
reproduced by a pulse with a finite magnitude. An example of the
scaled modulation functions and the corresponding magnetic field
for the generalized 3rd order UDD control are shown in
Fig.~\ref{Pulses}. Notice that due to the variable
transformation from $\theta$ to $t$, the magnetic field $B(t)$
does not have the symmetries as the scaled modulation functions
$f^{\pm}_N(\theta)$. For example, $B(t)$ is not periodic and the pulse
at different time has different width.

\textit{Summary} --
To summarize, we have proven that with $N$ ideal $\delta$-pulses for $\pi$-rotations,
the Uhrig dynamical decoupling
can suppress the pure dephasing or the longitudinal relaxation of a qubit
(or spin-1/2) coupled to an arbitrary bath, up to
$O\left(T^{N+1}\right)$. The qubit-bath coupling is not
eliminated to the same order as in generic concatenated dynamical decoupling.
But the remaining coupling would not induce the qubit decoherence
under consideration as it commutes with the relevant observable of the qubit.
As an extension of the proof, we also put forward a design of generalized UDD sequences
which are the ideal UDD $\delta$-pulse sequences superimposed with arbitrary
finite-duration pulses satisfying certain symmetry requirements. It should be pointed out
that the present proof of the UDD applies either to pure dephasing or longitudinal
relaxation and is limited to spin-1/2. It would be very interesting if the UDD can
be generalized for simultaneous suppression of transverse and longitudinal
relaxation and for higher spins.

This work was supported by Hong Kong RGC Project 2160285.
We would like to thank Jian-Liang Hu for discussions.


\end{document}